    \documentclass[usenatbib]{gh2013}
    
    \usepackage{mathptmx}       
    \usepackage{helvet}         
    \usepackage{courier}        
    \usepackage{makeidx}         
    \usepackage{graphicx}        
    \usepackage{multicol}        
    
    \usepackage{natbib}
    
\begin{document}

\title*{The LMT$^{\ast}$ Galaxies' 3 mm Spectroscopic Survey: First Results}
\titlerunning{The LMT Galaxies' 3 mm Survey}
\author{D. Rosa Gonz\'alez$^{1}$, P. Schloerb$^{2}$, 
O. Vega$^{1}$, L. Hunt$^{3}$, G. Narayanan$^{2}$,  
D. Calzetti$^{2}$,  M. Yun$^{2}$,  
E. Terlevich$^{1}$, R. J. Terlevich$^{1}$, Y. D. Mayya$^{1}$,
M. Ch\'avez$^{1}$, A. Monta\~na$^{1}$
and A. M. P\'erez Garc\'\i a$^{4}$
}
\authorrunning{D. Rosa Gonz\'alez et al.} 
\institute{
$^{\ast}$The Large Millimeter Telescope {\it Alfonso Serrano} is a binational project between
M\'exico and the United States of America. It is instaled in the Volc\'an Sierra
Negra, Puebla, M\'exico, and at the moment is operating using its 32m--diameter
reflecting primary surface. Further information about the project can be found
in http://www.lmtgtm.org/.
\\ 
$^{1}$INAOE, Luis Enrique Erro 1, Tonantzintla 72840, Puebla, M\'exico \\
\email{danrosa@inaoep.mx}\\ 
$^{2}$Department of Astronomy, University of Massachusetts, Amherst, MA 01003, USA\\
$^{3}$INAF - Osservatorio Astrofisico di Arcetri, Largo E. Fermi 5, 50125, Firenze, Italy \\
$^{4}$Instituto de Astrof\'\i sica de Canarias, C/ via L\'actea,  s/n, 38205, La Laguna, Tenerife, Spain\\
}
%
%
\maketitle


\vskip -2.5 cm  
\abstract{
The molecular phase of the interstellar medium (ISM) in galaxies offers
fundamental insight for understanding star-formation processes and how stellar
feedback affects the nuclear activity of certain galaxies.
We  present here Large Millimeter Telescope spectra obtained with the 
Redshift Search Receiver, a spectrograph that covers simultaneously 
the 3 mm band from 74 to 111 GHz with a spectral resolution of around 100 km\,s$^{-1}$.
Our selected galaxies, have been detected previously in HCN, and 
have different degrees of nuclear activity --- one normal galaxy (NGC~6946), 
the starburst prototype (M~82) and two ultraluminous infrared galaxies (ULIRGs, IRAS~17208-0014
and Mrk~231). 
We plotted our data in the HCO$^+$/HCN vs. HCN/$^{13}$CO diagnostic diagram 
finding that NGC~6946 and M~82 are located close to other normal galaxies; 
and that both IRAS~17208-0014 and Mrk~231 are close to the position of the well known 
ULIRG Arp~220 reported by~\citet{Snell2011}. 
We found that in Mrk~231 -- a galaxy with a well known active galactic
nucleus -- the HCO$^+$/HCN ratio is similar to the ratio observed in other
normal galaxies.
}
\section{Introduction}
The starburst-- active galactic nucleus (AGN)
connection is one of the most interesting and evolving 
topics of  modern astronomy. 
Learning about nuclear activity in galaxies implies segregating the different
components present in the nuclei of galaxies, plus allowing for a range of ages,
metal content and environmental differences. For example, when dealing with AGNs
and nuclear starbursts (SBs)  one has to take
into account that AGNs tend to appear preferentially in early type galaxies
while SBs tend to appear in Hubble types later than Sbc. 
This implies that -- on average -- physical parameters like 
metallicity or the age of the underlying stellar population are different in AGNs and SBs.
In general, Seyfert galaxies have metal content higher than the SB galaxies,  
thus the properties of the star formation in SB nuclei are not necessarily
similar to the star formation in Seyfert nuclei.
Part of the problem is that
galactic nuclei, where these processes take place, can be very dusty. This
compromises the interpretation of short-wavelength observations, and makes a 
multi-wavelength approach imperative.
One possible solution is to take advantage of sub-millimeter and millimeter (mm) spectra,
only recently possible thanks to new facilities.  
The mm-wavelength spectral surveys are beginning to provide sensitive measurements
of an astounding number of molecular transitions in starbursts and 
AGNs \citep[e.g.][]{Burillo2010,Aladro2011,Aladro2013}.
Such transitions allow the detailed study of
molecular chemistry in galactic nuclei, interpreted through theoretical models
only now starting to achieve the necessary sophistication 
\citep[e.g.][]{Papadopoulos2010,Kazandjian2012,Bayet2012}. 
The mm spectral line ratios provide diagnostics to interpret the relative contributions of different physical
environments in and around molecular clouds in the nuclear regions of galaxies.
Depending on the strengths of AGN accretion  and star formation, molecular
chemistry can be driven by Photo-Dissociated Regions (PDRs), Cosmic-Ray
Dominated Regions (CRDRs), X-ray Dominated Regions (XDRs), 
shock-dominated regions (or mechanically heated regions), and dense shielded
regions which lie deep in the molecular cloud ensembles. 
One of the main parameters that drives the evolution of a given galaxy is the 
star formation rate (SFR) which is related to the presence of dense shielded cores  
with n(H$_2$)$>3\times 10^4$ cm$^{-3}$. These regions are traced by molecules with high
dipole moment such as HCN and CS. In fact the HCN luminosity in CO bright 
galaxies seems to correlate very well with the infrared luminosity, a common tracer of the 
star formation activity \citep{Gao2004a,Gao2004b}.
Lines in the 3mm spectral region such as CO, HCN, HNC, HCO$^+$, and CS 
have been used 
\citep[e.g.][]{Kohno2001,Carpio2006,Snell2011}
to reveal and separate  PDRs and CRDRs,
associated with SF and intense starbursts, from XDRs, typically associated with 
AGNs 
\citep[e.g.][]{Papadopoulos2010,Meijerink2011}. 
It is interesting to see that the old problem of determining the source of energy 
in the nucleus of nearby galaxies -- star formation or accretion to black holes -- 
is still present in recent studies of molecular chemistry.
Rare isotopes such as C$^{18}$O and $^{13}$CO may trace rapid ISM enrichment
and in some cases they have been related to the presence of WR stars or Type II SNe 
(e.g., Costagliola et al. 2011).

\section{Observations and Data Reduction}
\label{sec:ObsRed}

The Redshift Search Receiver (RSR) is a dual polarization and dual beam instrument. The four broadband
receivers cover instantaneously the frequency range 74 -- 111 GHz
implying a very good relative calibration of the observed lines and making this instrument unique.  
Spectra are obtained by alternately positioning the source in one of the two
beams of the receiver.  In this way, the source is always being observed.  
Observations are made for 30 seconds in each beam and the spectra are then
subtracted from one another to obtain the spectrum of the difference between source and sky.
Calibration is done by the traditional chopper wheel method wherein a
measurement of an ambient load and sky are used to derive the system temperature and the antenna
temperature of sources according to the TA* scale. 
Typical values of the system temperature were around 100K.
Observations were always made by first peaking up the pointing and focus of the
antenna on a nearby radio source, having a pointing accuracy of
1''--2'' for the $\sim$25''~beam of the LMT at 3mm wavelength.
We  make use of the data analysis package called {\bf Dreampy} 
(Data Reduction and Analysis Methods in Python), written by G. Narayanan for 
reducing the data from the RSR. After converting and calibrating
the raw data and combining the frequency bands, the data in four pixels were 
averaged together using available routines in {\bf Dreampy} to produce final spectra for each galaxy.
\begin{figure*}[ht]
  \centering
  \includegraphics[width=13 cm]{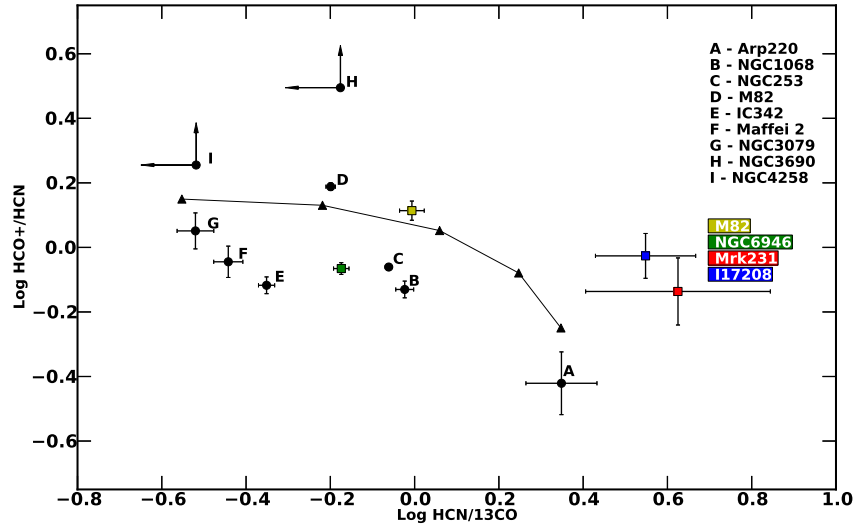}
  \caption{
The observed line ratios compared with different theoretical models. 
As an example, we show the comparison carried out by~\citet{Snell2011} in a sample of 
galaxies observed with the RSR mounted on the FCRAO telescope (black circles). 
The emission line ratios were modeled using the RADEX code \citep{Tak2007}.
The output of the models are marked with filled triangles connected by the solid
line. From the left to the right, the log--density of the different models are n(H$_2$)=4.0, 4.5,
5.0, 5.5 and 6.0 cm$^{-3}$. A temperature of 35~K was fixed for all the models.
In colored squares we mark the position of the galaxies observed by our team. 
}
\label{fig:TheFigure}
\end{figure*}
\section{Results and Conclusions}
\label{subsec:2}
The observations discussed in this paper include M~82 which 
was observed with the RSR at the LMT during the first-light scientific demonstration
observations phase in June 2011, and  
NGC~6946, IRAS~17208-0014 and Mrk~231 that where observed during the summer of 2013.
In the observed 3 mm spectra of these galaxies we detected several lines
including, in some cases, lines as weak as CN which traces very dense
molecular gas (n(H$_2)\sim 10^6$cm$^{-3}$).
However, we are reporting only our results on the HCN,  HCO$^+$ and $^{13}$CO lines.
In Fig.~\ref{fig:TheFigure} we plotted our results in the HCO$^+$/HCN vs. HCN/$^{13}$CO
diagnostic diagram. In that diagram the horizontal axis is an indicative of 
the dense gas fraction, where galaxies with high fraction of 
dense gas, N(H$_2)\sim 10^5$ cm$^{-3}$, are located to the right, 
and the vertical one is related with the presence of an AGN
where those galaxies with HCO$^+$/HCN much greater than 1 are better explained by
models that include X-ray dominated regions
~\citep[e.g.][]{Meijerink2011}. 
M~82 and NGC~6946 are close to other normal galaxies reported by~\citet{Snell2011}.  
In that paper they also observed M~82 (marked with a D in the Figure) 
with the same receiver but mounted in the FCRAO 15 m antenna. 
The differences between  the HCN, $^{13}$CO ratio can be explained by the different beam
sizes between the LMT and the FCRAO and the relative distribution of the dense gas traced by HCN 
and the more diffuse gas traced by $^{13}$CO. 
The two ultra--luminous galaxies (ULIRGs) observed,  IRAS~17208-0014 and Mrk~231, are close
to the position of the well known ULIRG Arp 220, indicating the presence of high 
amounts of dense gas in both galaxies.  
Notice that Mrk~231, a well known active galaxy classified as a Seyfert~1,  
has not an HCO$^+$ enhancement as predicted by the theoretical models.
By using all the available data of this galaxy, which cover all the
electromagnetic spectrum, we expect to better understand this discrepancy. 

We are witnessing the capability of the RSR--LMT system through  
the first results of an ongoing project where we will 
observe hundreds of galaxies with different degree
of central activity -- from HII galaxies, starbursts, LINERS, Seyfert 1--2,
and luminosities -- from normal galaxies to luminous and ultra luminous
infrared galaxies, and we will be able to relate with
high statistical significance the properties of nuclear activity with the complex molecular chemistry.

\begin{acknowledgement}
We thank the project head of LMT David Hughes and the INAOE director Alberto
Carrami\~nana for their continuous support to the project.
We also appreciate the work carried out by the LMT staff.

\end{acknowledgement}

\end{document}